\documentclass[aps,prl,reprint,superscriptaddress,showpacs,showkeys]{revtex4-1}

\usepackage{amsmath}
\usepackage[usenames]{color}
\usepackage{amssymb}
\usepackage{graphicx,epstopdf}

\DeclareMathOperator{\e}{e}
\DeclareMathOperator{\W}{W}
\DeclareMathOperator{\const}{const}

\begin{document}

\title{Scaling of Lyapunov exponents in chaotic delay systems}

\author{Thomas J\"ungling}
\email{thomas@ifisc.uib-csic.es}
\affiliation{Institute for Cross-Disciplinary Physics and Complex Systems, University of the Balearic Islands, 07122 Palma de Mallorca, Spain}

\author{Wolfgang Kinzel}
\affiliation{Institute for Theoretical Physics, University of W\"urzburg, Am Hubland, 97074 W\"urzburg, Germany}

\begin{abstract}
The scaling behavior of the maximal Lyapunov exponent in chaotic systems with time-delayed feedback is investigated. For large delay times it has been shown that the delay-dependence of the exponent allows a distinction between strong and weak chaos, which are the analogy to strong and weak instability of periodic orbits in a delay system. We find significant differences between scaling of exponents in periodic or chaotic systems. We show that chaotic scaling is related to fluctuations in the linearized equations of motion. A linear delay system including multiplicative noise shows the same properties as the deterministic chaotic systems.
\end{abstract}


\date{\today}

\maketitle

The cooperative behavior of nonlinear units is an active field of research, both from a fundamental point of view but also with applications in different scientific disciplines, from neurons to lasers~\cite{Arenas:08}. The nonlinear units interact by transmitting signals to their neighbors. Often the transmission time is longer than the internal time scales of these units; the coupling has a delay time. Dynamical systems with time-delayed couplings may lead to high-dimensional chaos, and networks of such units may synchronize to clusters of common chaotic trajectories~\cite{Lakshmanan:11,*Erneux:09}.
Chaos is characterized by the maximal Lyapunov exponent of the network which measures the sensitivity to initial conditions. In this Letter we study a fundamental aspect of dynamical systems with time-delayed feedback, namely the scaling of the Lyapunov exponent with the delay time. We find that the scaling behavior of chaotic systems shows anomalies compared to the corresponding scaling of periodic systems with time-delayed couplings. These anomalies can be related to linear networks with time-delayed couplings and multiplicative noise. We consider a nonlinear dynamical system defined by the equations of motion
\begin{equation}
\dot{\mathbf{x}}=\mathbf{f}(\mathbf{x})+\mathbf{K}\cdot\mathbf{x}_\tau\;,
\label{system}
\end{equation}
where $\mathbf{x}\equiv\mathbf{x}(t)\in\mathbb{R}^N$ and $\mathbf{x}_\tau\equiv\mathbf{x}(t-\tau)$, with the delay time $\tau>0$. For simplicity we choose linear coupling described by the matrix $\mathbf{K}\in\mathbb{R}^{N\times N}$. The vector field $\mathbf{f}:\mathbb{R}^N\rightarrow\mathbb{R}^N$ can be an arbitrary but smooth and differentiable nonlinear function. Our model also includes large complex systems, i.e., the function $\mathbf{f}(\cdot)$ can describe many nonlinear dynamical nodes of a network, each of them with several degrees of freedom. In such a delay-coupled network the matrix $\mathbf{K}$ would correspond to the interaction strengths between the nodes. We are interested in Lyapunov exponents of the system described by Eq.~(\ref{system}). A Lyapunov exponent is a measure for the evolution of a small perturbation, which is calculated by linearizing Eq.~(\ref{system})
\begin{equation}
\dot{\delta\mathbf{x}}=\mathbf{Df}(\mathbf{x})\cdot\delta\mathbf{x}+\mathbf{K}\cdot\delta\mathbf{x}_\tau\;.
\label{linsys}
\end{equation}
Here $\mathbf{Df}(\mathbf{x})$ denotes the Jacobian of $\mathbf{f}(\cdot)$ with $(\mathbf{Df}(\mathbf{x}))_{ij}=\partial f_i/\partial x_j|_{\mathbf{x}(t)}$. It is evaluated at the trajectory $\mathbf{x}(t)$ and is therefore a time-dependent matrix. In presence of a chaotic trajectory $\mathbf{x}(t)$ the matrix elements are non-periodic. The Lyapunov exponent is defined from the evolution of the linear system Eq.~(\ref{linsys}) with typical initial conditions
\begin{equation}
\lambda=\lim_{t\rightarrow\infty}\frac{1}{t-t_0}\ln\frac{\|\delta\mathbf{x}(t)\|}{\|\delta\mathbf{x}(t_0)\|}\;.
\end{equation}
For the following discussion, $\lambda$ denotes the maximum Lyapunov exponent on a typical chaotic attractor.

\paragraph{Strong and weak chaos.}

For sufficiently large delay $\tau\gg\tau_0$, where $\tau_0$ is system-dependent, recently it has been shown, that the maximum Lyapunov exponent $\lambda$ as a function of the delay time shows two major types of scaling called strong or weak chaos~\cite{Heiligenthal:11}. In strong chaos, the Lyapunov exponent reaches a limit value
\begin{equation}
\lim_{\tau\rightarrow\infty}\lambda(\tau)=\lambda_0\;.
\label{stronglimit}
\end{equation}
In weak chaos, $\lambda$ decreases towards zero in the same limit. But it scales with the delay time as $\lambda\propto\tau^{-1}$, such that
\begin{equation}
\lim_{\tau\rightarrow\infty}\tau\cdot\lambda(\tau)=\hat{\mu}\;.
\label{weaklimit}
\end{equation}
We call the product $\lambda\tau$ the \textit{delay-normalized} Lyapunov exponent. The scaling of $\lambda$, by which we distinguish between strong and weak chaos, depends on the sign of an auxiliary exponent $\lambda_0$. This exponent is given by the partial linearization of Eq.~(\ref{system}), in which the delayed feedback is omitted
\begin{equation}
\dot{\delta\mathbf{x}}_0=\mathbf{Df}(\mathbf{x})\cdot\delta\mathbf{x}_0\;.
\label{linsys0}
\end{equation}
Note that, however, the full trajectory $\mathbf{x}(t)$ of the delay system~(\ref{system}) enters both linearizations Eq.~(\ref{linsys}) and Eq.~(\ref{linsys0}). The auxiliary exponent then reads
\begin{equation}
\lambda_0=\lim_{t\rightarrow\infty}\frac{1}{t-t_0}\ln\frac{\|\delta\mathbf{x}_0(t)\|}{\|\delta\mathbf{x}_0(t_0)\|}\;.
\label{lambda0def}
\end{equation}
In the following we call it the \textit{sub-exponent}, because it is a special conditional exponent describing a subsystem of the original system~\cite{Lepri:94}. If $\lambda_0>0$, there is strong chaos and $\lambda_0$ from Eq.~(\ref{stronglimit}) and Eq.~(\ref{lambda0def}) coincide. Otherwise, if $\lambda_0<0$, weak chaos is present and the limit $\hat{\mu}$ from Eq.~(\ref{weaklimit}) does not depend trivially on $\lambda_0$ like in strong chaos~\cite{Heiligenthal:11}.

\paragraph{Periodic dynamics}

The delay system Eq.~(\ref{system}) may have unstable periodic solutions $\mathbf{x}(t)=\mathbf{x}(t+T_p)$ with period $T_p=\tau/n$, $n\in\mathbb{N}$, including fixed points $\mathbf{x}(t)\equiv\mathbf{x}^*$. Such a periodic orbit reappears periodically, when the delay time is varied~\cite{Yanchuk:09}. So it is possible to define and observe scaling laws of its Lyapunov exponent, which is the real part of the so-called Floquet exponent for orbits. In analogy to strong and weak chaos, the orbit would be called strongly or weakly unstable, if its sub-exponent is positive or negative, respectively.
We show three scaling laws for the Lyapunov exponent of periodic orbits, which in case of diagonal coupling $\mathbf{K}=k\boldsymbol{1}$ can be derived analytically. These laws remain valid for arbitrary coupling. Eq.~(\ref{linsys}) can be transformed into a system with only constant coefficients $\{A_{i,j}\}$ by the Floquet-ansatz $\delta\mathbf{z}(t)=\mathbf{Q}(t)\cdot\delta\mathbf{x}(t)$, where $\mathbf{Q}(t)=\mathbf{Q}(t+\tau)$ is a suitable transformation matrix~\cite{Hale:93,*Just:2000}. The resulting system
\begin{equation}
\dot{\delta\mathbf{z}}=\mathbf{A}\cdot\delta\mathbf{z}+k\cdot\delta\mathbf{z}_\tau
\end{equation}
provides a characteristic equation, which reveals the maximum Lyapunov exponent
\begin{equation}
\lambda=\Re\left\{\lambda_0+\frac{1}{\tau}\W\left(k\tau\e^{-(\lambda_0+i\omega_0)\tau}\right)\right\}\;.
\label{lamberteq}
\end{equation}
Here $\lambda_0$ is the sub-exponent, which is the maximum real part of the eigenvalues of $\mathbf{A}$. $\W:\mathbb{C}\rightarrow\mathbb{C}$ is the Lambert function with $\W(z)\exp(\W(z))=z$ for $z\in\mathbb{C}$. The imaginary part $\omega_0$ can be omitted for large delay times~\cite{Flunkert:10}. From the general expression Eq.~(\ref{lamberteq}), we derive three limiting expressions. For strong instability $\lambda_0>0$ and large delay times it reads
\begin{equation}
\lambda=\lambda_0+k\e^{-\lambda_0\tau}\;,
\label{perstrong}
\end{equation}
meaning that the difference $\lambda-\lambda_0$ vanishes exponentially with increasing $\tau$. For weak instability $\lambda_0<0$ the limit of the delay-normalized exponent $\lim\limits_{\tau\rightarrow\infty}\tau\lambda(\tau)=\hat{\mu}$ becomes
\begin{equation}
\hat{\mu}=\ln\left(-\frac{k}{\lambda_0}\right)\;.
\label{perweak}
\end{equation}
When a system parameter in Eq.~(\ref{system}) is changed, $\lambda_0$ may cross zero and one observes a transition from weak to strong instability. Eq.~(\ref{perweak}) shows, that $\hat{\mu}$ diverges logarithmically with $\lambda_0$, when the transition point is approached. Finally, at the critical point $\lambda_0=0$ the delay-normalized exponent scales with respect to $\tau$ as
\begin{equation}
\lambda\tau=\W(k\tau)\;.
\label{percritical}
\end{equation}

\paragraph{Anomalous scaling.}

\begin{figure}
\includegraphics[width=\columnwidth]{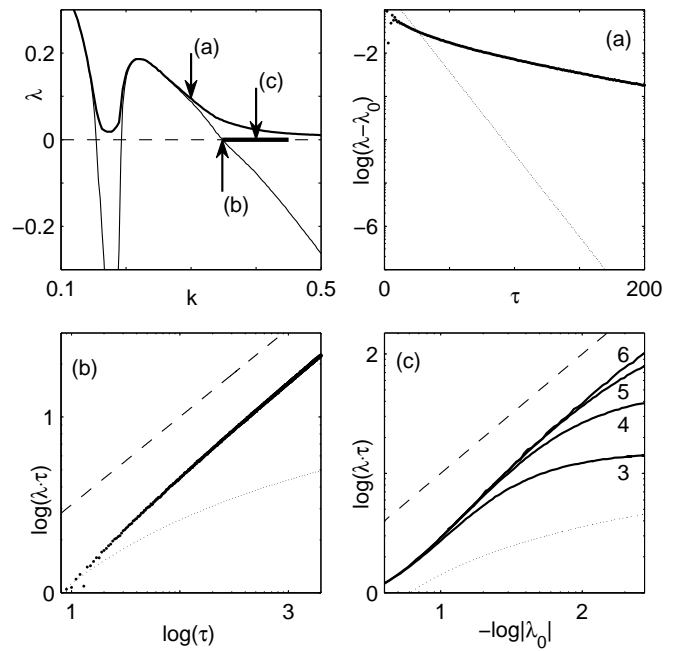}
\caption{Lyapunov exponents for logistic map with delayed feedback. Top left: $\lambda_0(k)$ (lower curve, thin) and $\lambda(k)$ (upper curve, thick) for $\tau=100$ with transitions between strong and weak chaos. Bar at arrow (c) indicates selected interval in figure (c). (a): Convergence of $\lambda(\tau)$ towards $\lambda_0$ in strong chaos (big dots), compared with the curve $k\cdot\exp(-\lambda_0\tau)$ (small dots). (b): $\lambda(\tau)\cdot\tau$ at critical point $\lambda_0=0$ (big dots), compared with $\W(k\tau)$ (small dots) and $\sqrt{\tau}$ (dashed). (c): $\lambda\tau(\lambda_0)$ in weak chaos (solid lines) for $\tau=10^n,n\in\{3,4,5,6\}$, compared with $\ln(-k/\lambda_0)$ (small dots) and $1/|\lambda_0|$ (dashed).}
\label{logisticscaling}
\end{figure}

Returning to chaotic dynamics, we observe a significant deviation from each of the three scaling laws presented above. These anomalies are present in every chaotic system with time-delayed self-feedback, which we have studied, including the R\"ossler, Lorenz and Lang-Kobayashi model as well as the H\'enon map, logistic map and skew tent map. We conclude that this behavior is generic. Exceptions are the Bernoulli map and the continuous Ikeda system, where the drive terms are constants. We exemplify the anomalous scaling by means of the logistic map $M(x)=4x(1-x)$ with time-delayed feedback
\begin{equation}
x_{t+1}=(1-k)M(x_t)+kM(x_{t-\tau})\;,
\label{discretemap}
\end{equation}
where $t\in\mathbb{Z}$ and $\tau\in\mathbb{N}$ are normalized by the time step of the map. The scaling laws as presented in the following are the same for continuous flows and for discrete maps, except for insignificant prefactors and additive constants. From all systems mentioned above, the logistic map shows the scaling laws in the clearest way.

For strong chaos ($\lambda_0>0$), the maximum exponent $\lambda$ converges to $\lambda_0$, if the delay time approaches infinity, and indeed we find an exponential convergence, as shown in Fig.~\ref{logisticscaling}a for the logistic map. But the characteristic decay constant is much closer to zero than expected from the viewpoint of the previous Floquet analysis, meaning that $\lambda(\tau)$ is generally larger than the real part of a comparable Floquet exponent and decays slower to its limit,
\begin{equation}
\lambda-\lambda_0\propto\e^{-\alpha\tau}\;,
\end{equation}
with $0<\alpha<\lambda_0$.

The second anomaly appears in the weak chaos regime ($\lambda_0<0$) with respect to variation in $\lambda_0$. The delay-normalized exponent $\lambda\tau$ depends on $\lambda_0$. In contrast to the logarithmic divergence in Eq.~(\ref{perweak}), for chaotic dynamics we observe a power law
\begin{equation}
\hat{\mu}\propto\lambda_0^{-1}\;,
\label{lambda0to-1}
\end{equation}
which is clearly exemplified by the logistic map with time-delayed feedback, Fig~\ref{logisticscaling}c. For finite delay times, however, this power law is incomplete, but by increasing the delay time one observes that the curves converge to the prediction Eq.~(\ref{lambda0to-1}).

The most outstanding anomalous scaling behavior occurs at the transition between strong and weak chaos, where $\lambda_0=0$. Instead of the slow growth described by Eq.~(\ref{percritical}), for large delays the normalized exponent obeys a power law
\begin{equation}
\lambda\tau\propto\sqrt{\tau}
\label{criticalsqrtlaw}
\end{equation}
as shown in Fig.~\ref{logisticscaling}b. In the vicinity of this critical point one finds a crossover between the divergence Eq.~(\ref{criticalsqrtlaw}) and the saturation Eq.~(\ref{weaklimit}). In the following we show that the anomalies are connected to the strength of the fluctuations in the driving term $\mathbf{Df}(\mathbf{x}(t))$ of Eq.~(\ref{linsys}).

\paragraph{Stochastic model.}

Which is the simplest model displaying anomalous scaling? If we replace the fluctuations from the chaotic trajectory by noise in Eq.~(\ref{linsys}), a simple model is a one-dimensional linear delay system with multiplicative noise
\begin{equation}
\dot{z}=(\lambda_0+\eta(t))z+\kappa z_\tau\;.
\label{smodel}
\end{equation}
The variable $z>0$ can be understood as a correspondence to $\|\delta\mathbf{x}\|$ in a real system. $\lambda_0$ can be identified with the sub-exponent and $\kappa$ replaces the feedback gain. The term $\eta(t)$ introduces fluctuations with $\langle\eta(t)\rangle=0$, which model the non-periodic time-dependence of the coefficients $\mathbf{Df}(\mathbf{x}(t))$. $\eta(t)$ should most naturally be given by an Ornstein-Uhlenbeck process with correlation time $T_c$. But since we investigate the large delay regime with a timescale separation $\tau\gg T_c$, we can replace the process by white noise, $\eta(t)=\sqrt{2D}\xi(t)$ and $\langle\xi(t)\xi(t+t')\rangle=\delta(t')$, where $D$ is the diffusion constant of the process. The stochastic delay-differential equation is interpreted in the sense of Stratonovich, in order to guarantee, that an originally smooth process with finite correlation times is modeled. In this interpretation we can transform Eq.~(\ref{smodel}) by $w=\ln(z)$, which emerged to be very useful for analytical discussion and also for numerical integration. Then the logarithm $w$ obeys an equation with additive noise
\begin{equation}
\dot{w}=\lambda_0+\sqrt{2D}\xi(t)+\kappa\e^{w_\tau-w}\;.
\label{logmodel}
\end{equation}
This equation reveals the essential nonlinear character of the seemingly linear system Eq.~(\ref{smodel}). In the absence of noise ($D=0$) we can directly calculate the Lyapunov exponent of this model via an exponential ansatz $z=z_0\exp(\lambda t)$ or $w=w_0+\lambda t$, and we obtain the same expression as Eq.~(\ref{lamberteq}) for $\omega_0=0$ and $\kappa=k$. For the general case with noise $D>0$, we were not able to derive a closed solution for the Lyapunov exponent, which in this case would be identical with the drift of the logarithm $w$. The main problem appears in the formulation of a corresponding Fokker-Planck equation for the stochastic delay differential equations. The so-called \textit{conditional average drift}, which describes the joint probability distribution $P(w|w_\tau)$ remains generally unknown~\cite{Guillouzic:99}. Nevertheless, numerical solutions of Eq.~(\ref{logmodel}) verify all three anomalies, which we found for the chaotic systems. This result means, that the introduction of fluctuations changes the scaling laws qualitatively, from those of periodic dynamics to those of chaotic dynamics.

By means of appropriate approximations, we were able to derive analytical limit expressions. Regarding the first anomaly in the strong chaos regime, $\lambda_0>0$, we obtain for $D<\lambda_0$
\begin{equation}
\lambda\approx\lambda_0+\kappa\e^{(D-\lambda_0)\tau}\;.
\end{equation}
It explains the slower convergence of $\lambda$ towards $\lambda_0$ directly by means of the multiplicative noise intensity $D$. In the parameter regime corresponding to weak chaos, $\lambda_0<0$, we could derive a lower bound for the exponent, such that the limit expression for the delay-normalized exponent becomes
\begin{equation}
\hat{\mu}(\lambda_0,\kappa,D)\ge\ln\left(\frac{\kappa}{D}\right)-\psi\left(-\frac{\lambda_0}{D}\right)\;.
\end{equation}
Here $\psi(z)$ is the digamma function with $\psi(z)=d/dz(\ln\Gamma(z))$, $\Gamma(z+1)=z\Gamma(z)$. This formula incorporates both limits for the almost periodic case and the case of strong fluctuations. The limit of noise-free dynamics is $D\ll\lambda_0$ and reveals $\hat{\mu}=\ln|\kappa/\lambda_0|$, which is the expected result. In the case of strong noise or $\lambda_0\rightarrow0-$, it is in leading order $\hat{\mu}\ge-D/\lambda_0$, which is the power law observed in chaotic dynamics.
Considering the last and most prominent anomaly, which occurs at the critical point $\lambda_0=0$, we approximate the dynamics in $w$ by a random-walk with a reflecting boundary at $w_\tau$. The result is a one-sided diffusion, which results in a drift with the delay-normalized exponent
\begin{equation}
\lambda\tau\propto\sqrt{D\tau}\;.
\end{equation}
These results agree with our numerical simulations of various chaotic systems mentioned above.

\paragraph{From noise-free to noisy.}

\begin{figure}
\includegraphics[width=\columnwidth]{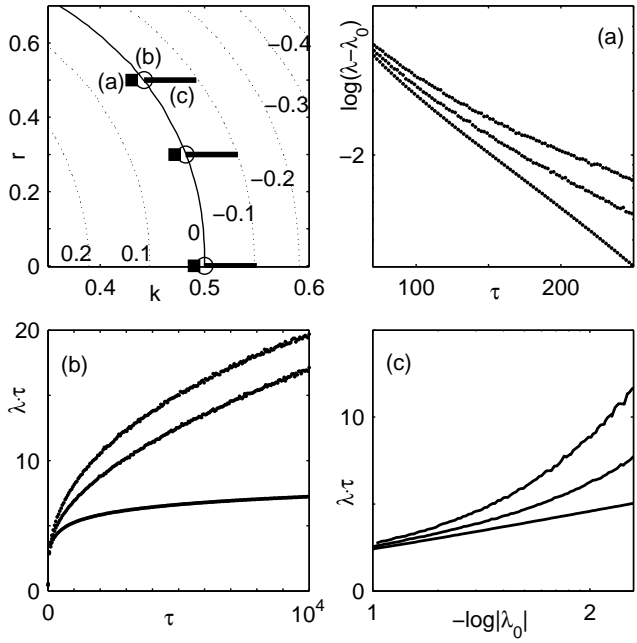}
\caption{Gradual generation of anomalous scaling of maximal Lyapunov exponents in the skew Bernoulli map. Top left: Parameter plane of $(k,r)$ with lines of constant $\lambda_0$ as labeled. Squares, circles and thick solid lines mark the parameter values examined in (a)-(c), respectively. Lowest curve in each of fig.~(a)-(c) corresponds to $r=0$, with no fluctuations in the linearized equations, middle curve to $r=0.3$ and upper curve to $r=0.5$. (a): Strong chaos $\lambda_0>0$ (close to transition point), fluctuations decrease the decay constant $\alpha$ of $\lambda-\lambda_0\propto\exp(-\alpha\tau)$. (b): Critical point $\lambda_0=0$, fluctuations change scaling from $\lambda\tau\propto\W(\tau)$ to $\lambda\tau\propto\sqrt{\tau}$. (c): Weak chaos $\lambda_0<0$ at $\tau=1.5\cdot10^4$. Fluctuations change scaling from $\hat{\mu}\propto\ln|k/\lambda_0|$ towards $\hat{\mu}\propto\lambda_0^{-1}$ for $\lambda_0\rightarrow0-$.}
\label{bernoulliscaling}
\end{figure}
Finally, we introduce a simple chaotic model, where we can tune the fluctuations of the coefficients in its linearized equations by changing a parameter $r$. Thus, we observe a gradual change of the three different scaling behaviors discussed before. We consider the \textit{skew Bernoulli map}
\begin{equation}
M_r(x)=
\begin{cases}
\frac{2}{1+r}x & x\le \tfrac{1}{2}(1+r) \\
\frac{2}{1-r}(x-\tfrac{1}{2}(1+r)) & x>\tfrac{1}{2}(1+r)
\end{cases}
\end{equation}
with delayed feedback as in Eq.~(\ref{discretemap}). The map has two different slopes, namely $m_1=2/(1+r)$ in the left regime and $m_2=2/(1-r)$ in the right regime. The linearization for the Lyapunov exponent $\lambda$ reads
\begin{equation}
\delta x_{t+1}=(1-k)M_r'(x_t)\delta x_t+kM_r'(x_{t-\tau})\delta x_{t-\tau}\;,
\label{bernilin}
\end{equation}
and the sub-exponent $\lambda_0$ is determined by the partial linear system, in which the delay term has been omitted. The parameter $r$ allows us to change the degree of asymmetry in the map. For $r=0$ the map is identical to the original Bernoulli map with constant slope $M_0'(x_n)\equiv2$. This leads to $\lambda_0=\ln(1-k)+\ln2$, and due to the constant slope no fluctuations from the chaotic trajectory enter the linearization Eq.~(\ref{bernilin}). The Lyapunov exponent $\lambda$ can be calculated analytically, and for large delays $\tau\gg1$ we obtain the same scaling behavior as for the case of periodic orbits or steady states as described above. Increasing the parameter $r$ gradually introduces the fluctuations of the chaotic system into the linearization by the two different slopes $m_1$ and $m_2$. In order to study this effect of multiplicative noise systematically, we aim to increase the parameter $r$ starting at $r=0$, while setting the sub-exponent $\lambda_0$ to any desired value. To this end, we have first recorded a phase diagram $\lambda_0(k,r)$ for a sufficiently large delay. There exist parameterizable curves connecting $r(p)$ and $k(p)$, such that $\lambda_0(k(p),r(p))=\const$. For a fixed value of $\lambda_0>0$, we scan the delay dependence of $\lambda$ for different values of $r$, and we clearly observe the emergence of the first anomaly, as chaotic fluctuations enter the linear system. The other two anomalies can be demonstrated in an analog way by changing from $r=0$ to $r\neq0$. Again, most significant is the impact of multiplicative noise at the critical point $\lambda_0=0$.

In summary, strong and weak chaos is related to the fluctuations of the coefficients in linear equations defining Lyapunov exponents. In particular, at the transition from strong to weak chaos, these fluctuations lead to scaling laws which are different from corresponding ones of periodic systems. The scaling of chaotic systems is reproduced by linear systems with multiplicative noise and time-delayed feedback.

\end{document}